\documentclass[amsmath,amssymb,aps,prl,twocolumn,showpacs,superscriptaddress]{revtex4-1}
\usepackage{graphicx}
\usepackage{dcolumn}
\usepackage{bm}
\usepackage{xr}
\makeatletter
\usepackage{hyperref}
\usepackage{xcolor}
\hypersetup{
  colorlinks   = true,
  urlcolor     = black,
  linkcolor    = blue,
  citecolor   = blue
}

\begin{document}

\title{Depletion attraction favors the elastic response of emulsions flowing in a constriction}

\author{I. Golovkova}
\affiliation{Laboratoire Jean Perrin, Institut de Biologie Paris-Seine (IBPS), CNRS UMR 8237, Sorbonne Universit\'e, 4 place Jussieu, 75005 Paris, France}
\author{L. Montel}
\affiliation{Laboratoire Jean Perrin, Institut de Biologie Paris-Seine (IBPS), CNRS UMR 8237, Sorbonne Universit\'e, 4 place Jussieu, 75005 Paris, France}
\author{E. Wandersman}
\affiliation{Laboratoire Jean Perrin, Institut de Biologie Paris-Seine (IBPS), CNRS UMR 8237, Sorbonne Universit\'e, 4 place Jussieu, 75005 Paris, France}
\author{T. Bertrand}
\email[]{t.bertrand@imperial.ac.uk}
\affiliation{Department of Mathematics, Imperial College London, South Kensington Campus, London SW7 2AZ, England, UK}
\author{A. M. Prevost}
\affiliation{Laboratoire Jean Perrin, Institut de Biologie Paris-Seine (IBPS), CNRS UMR 8237, Sorbonne Universit\'e, 4 place Jussieu, 75005 Paris, France}
\author{L.-L. Pontani}
\email[]{lea-laetitia.pontani@sorbonne-universite.fr}
\affiliation{Laboratoire Jean Perrin, Institut de Biologie Paris-Seine (IBPS), CNRS UMR 8237, Sorbonne Universit\'e, 4 place Jussieu, 75005 Paris, France}
\date{\today}

\begin{abstract}
We study the elasto-plastic behavior of dense attractive emulsions under mechanical perturbation. The attraction is introduced through non-specific depletion interactions between the droplets and is controlled by changing the concentration of surfactant micelles in the continuous phase. We find that such attractive forces are not sufficient to induce any measurable modification on the scalings between the local packing fraction and the deformation of the droplets. However, when the emulsions are flown through 2D microfluidic constrictions, we uncover a measurable effect of attraction on their elasto-plastic response. Indeed, we measure higher levels of deformation inside the constriction for attractive droplets. In addition, we show that these measurements correlate with droplet rearrangements that are spatially delayed in the constriction for higher attraction forces.
\end{abstract}


\maketitle

\section{INTRODUCTION}
The flow of particulate systems is a problem of great importance both theoretically and practically, with direct applications to the industry. It is relevant for a wide range of soft materials, from granular packings to foams and emulsions. While these materials present obvious differences, they share universal features, e.g. they generically undergo what is known as a jamming transition \citep{Liu2010,Vanhecke2009}. As the particle or droplet volume fraction $\phi$ increases, this rigidity transition between liquid and amorphous-solid states controls the phase behavior of these disordered solids. At a critical volume fraction $\phi_c$ (random close packing), the system jams and develops a yield stress \citep{Cates1998,Ohern2003,Olsson2007,Goyon2008}. The mechanical and rheological properties, such as the elastic modulus or the local pressure, of these systems are known to display a power law dependence with the distance to the jamming onset $(\phi-\phi_c)$ \citep{Ohern2002,Ohern2003,Olsson2007,Mason1995,Lacasse1996,Ellenbroek2006,Majmudar2007,Jorjadze2013}.\\
\indent Jammed solids are characterised by a spatially heterogeneous network of interparticle contacts, with a broad distribution of contact forces exhibiting an exponential tail \citep{Jaeger1996,Ohern2003,Majmudar2007,Brujic2003} in which only a small subset of the particles sustain most of the mechanical load \citep{Liu1995,Cates1999,Majmudar2005,Zhou2006}. Below the yield stress, these systems responds elastically, while above it, they deform and flow plastically \citep{Chen2010}. In these soft glassy flows, it was shown that stress and strain rates are coupled nonlocally \citep{Goyon2008,Kamrin2012,Bocquet2009}. 
In two-dimensional materials, the flow properties can easily be probed both at the microscopic and macroscopic scales \cite{Desmond2015,Chen2015,Hartley2003,Lauridsen2004,Kabla2007,Dollet2007,Utter2008,Keim2014,Keim2015,Bares2017}. As a consequence, previous experimental studies examined the microscopic rearrangements in a variety of two-dimensional model systems under stress \citep{Lauridsen2002, Gai2016}.
This plastic flow is generically governed by local structural rearrangements which relieve stresses and dissipate energy \citep{Argon1979,Goyon2008,Desmond2015,Chen2015}. Local plastic rearrangements have been connected to the fluctuating macroscopic flow in both simulations \citep{Durian1995,Kabla2003,Maloney2004,Maloney2006,Mansard2013} and theoretical studies \citep{Falk1998,Picard2004,Goyon2008,Bocquet2009,Kamrin2012} of model systems. Nevertheless, the intimate link between the microscopic dynamics of an amorphous material and its macroscopic elasto-plastic response is still an open question for a broad class of more realistic materials.\\
\indent In emulsions, the use of surfactants prevents the coalescence of the droplets and leads to short-range purely repulsive droplet-droplet interactions \citep{Desmond2013,Desmond2015,Chen2015}. As such, dense stable emulsions are examples of jammed solids. In the last decades, a number of experimental works studied the structural, mechanical and rheological properties of purely repulsive emulsions \citep{Desmond2013,Hebraud1997,Coussot2002,Becu2006,Jop2012,Jorjadze2013, Lin2016}. 
In particular, as in other soft materials \citep{Lundberg2008,Graner2008,Cohen-Addad2013,Marmottant2008,Keim2014,Bi2015,Bares2017}, recent studies in quasi-2D flowing emulsions have also highlighted the importance of T1 events for local rearrangements and stress redistribution~\citep{Chen2015,Desmond2015}. 
Monodisperse emulsions allow one to study material properties such as grain boundaries, dislocations and plasticity \citep{Bragg1947,Schall2004,Schall2007,Arciniaga2011,Arif2012}; in particular, a recent study showed the existence of a spatiotemporal periodicity in the dislocation dynamics of these emulsions \citep{Gai2016}.
However, none of these studies have so far adressed the question of how interdroplet attractive forces modify the flow response of these emulsions.\\
\indent Indeed, in a variety of natural settings and industrial applications, emulsion droplets do display additional attractive interactions that have been shown to change the nature of the jamming transition \citep{Trappe2001,Lois2008, Jorjadze2011}. In contrast with the purely repulsive case, droplets in attractive emulsions can form bonds and thus a soft gel-like elastic structure which can sustain shear stresses below isostaticity \citep{Bibette1993,Poulin1999,Becu2006,Datta2011}. However, the microscopic dynamics of the material, i.e. at the scale of the particles, was not explored. As a consequence, it is of particular importance to ask how the response to stress and in particular, the structural and mechanical properties of emulsions are modified by the presence of attractive interactions. Despite their broad applicability, our understanding of the influence of particle-particle interactions on the macroscopic properties of soft matter systems with attractive interactions is currently hindered by a crucial lack of controlled experimental settings. \\
\indent In this article, we propose a first step towards completing our understanding of the microscopic origin for the macroscopic properties of adhesive emulsions. In particular, we study emulsions in which droplets interact through depletion attraction. First, we find that the static structure of 2D polydisperse emulsions remains unchanged by the introduction of depletion forces. However, the response of 2D monodisperse emulsions under mechanical constraint is impacted by the presence of depletion forces. Indeed, we flow the droplets through a microfluidic constriction in which they have to undergo elasto-plastic remodelling in order to go from a wide channel to a narrow one. In particular, we find that attractive droplets deform more inside the constriction, which we correlate to a shift in the expected position of rearrangements. These findings show that depletion attraction forces are sufficient to modify the elasto-plastic response of dense emulsions under a mechanical perturbation. This attraction, even though it is not evidenced in static conditions, impairs rearrangements and in turn promotes an enhanced elastic response under flow.

\section{MATERIALS AND METHODS}
\subsection{Emulsion preparation}
\indent For static experiments, polydisperse emulsions are prepared using a pressure emulsifier (Internal Pressure Type, SPG Technology co.). Silicon oil (viscosity $50 \mathrm{mPa.s}$, Sigma Aldrich) is pushed through a cylindrical Shirasu Porous Glass membrane decorated with $10~\mu \mathrm{m}$ pores, direcly into a 10mM SDS solution that is maintained under vigorous agitation. The resulting droplets display an average diameter of $42~\mu \mathrm{m}$ (polydispersity $21\%$). In order to prepare the emulsion with both SDS concentrations, we use the same droplets and only replace their continuous phase. To do so, the emulsion is washed in a separating funnel in order to replace the continuous phase by solutions of 10 or 45mM SDS in a water/glycerol mixture (60:40 in volume). This enhances the optical quality of the oil/water interface visualization through bright field and confocal microscopy.\\
\indent For experiments in the constriction, we use monodisperse emulsions with an average droplet diameter of $45~\mu\mathrm{m}$ (polydispersity $3.9\%$). These emulsions are obtained with a custom made flow-focusing microfluidic set-up (channel size $60~\mu \mathrm{m} \times 60~\mu \mathrm{m}$, width at the flow-focusing junction $30~\mu \mathrm{m}$). We use the same oil and continuous phases for polydisperse and monodisperse emulsions.

\subsection{Observation and image analysis of 2D static packings}
\begin{figure*}
	\centering
		\includegraphics[height=6.2cm]{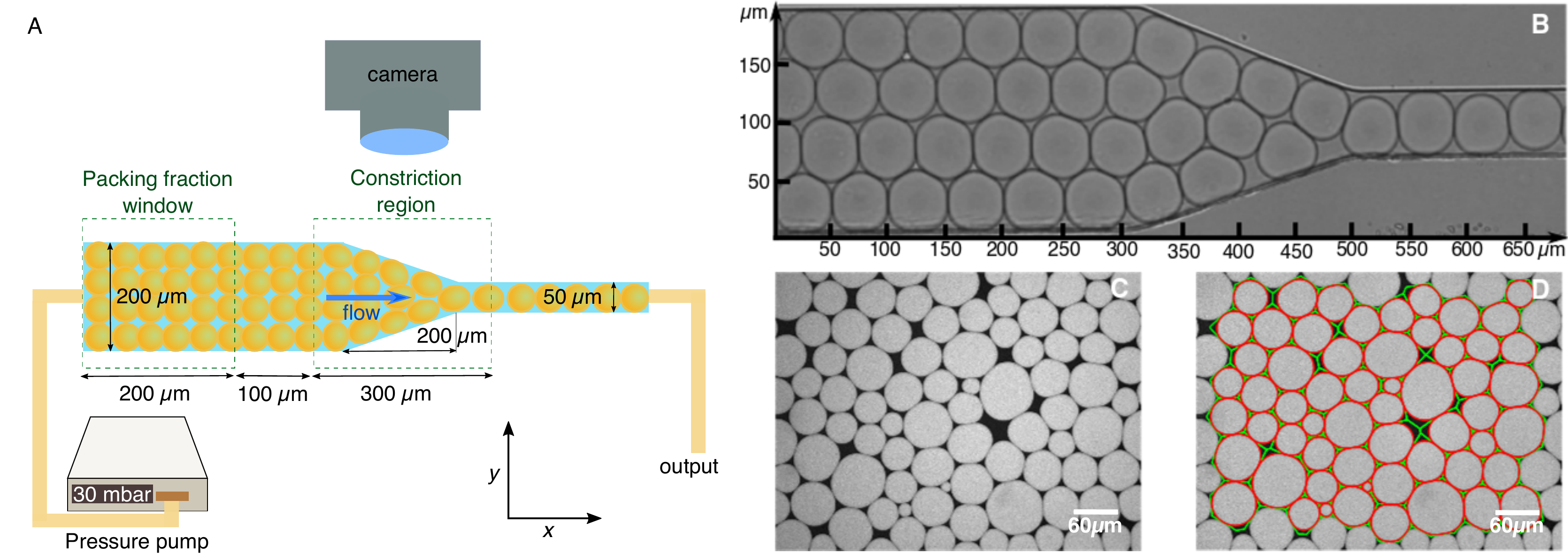}
	\caption{{Experimental set-up and image analysis} --- {(A)} The oil in water emulsion is pushed using a pressure pump (P = 30 mbar) through the microfluidic channel that consists of three parts: a 200~$\mu$m wide channel, a constriction, and 50~$\mu$m wide channel. The depth of the channel is 50~$\mu$m over the whole length, and the diameter of the droplets is $\approx 45~\mu$m. {(B)} Typical image of a monodisperse emulsion flowing in the constriction. In the area of the constriction, the flow of the droplets is imaged in bright field microscopy at 20 fps. The packing fraction of the emulsion is determined within the window of 200x200~$\mu$m located before the constriction area. {(C)} A typical confocal microscopy image of compressed 2D droplets at [SDS]=10mM. {(D)} Result of the image analysis performed on (C). Droplet contours are shown in red and watershed tessellation cells with the green curves. Based on these measurements, we calculate the local packing fraction $\phi_l$ as the ratio between the area of the droplet and that of its corresponding watershed tesselation cell, as well as the shape parameter $\mathcal{A}$. }
\label{Fig1}
\end{figure*}

When studying 2D static packings, we consider polydisperse emulsions that are fluorescently labelled with Nile Red (Sigma Aldrich). To label the emulsion, we incubate it overnight in a SDS buffer (with [SDS] = 10 or 45 mM) saturated in Nile Red allowing the dye to partition between the oil and water phases over time. A  $10~\mu\mathrm{L}$ drop of emulsion is placed between a microscope glass slide (76 x 26 mm, Objekttrager) and a cover slip (24 x 60mm, Knittel Glaser) separated by 50~$\mu$m spacers (polymethylmethacrilate -PMMA- film, Goodfellow). Droplets are imaged through confocal microscopy (Spinning Disc Xlight V2, Gataca systems) using a 20x objective.\\
\indent To study the local structure of these static packings, we use a custom Matlab (MathWorks) routine that works as follows. We first threshold the images and perform a watershed tessellation, we then measure the perimeter $p$ and area $a$ of each droplet as well as the area $a_{c}$ of the associated watershed tesselation cell (see Fig.~\ref{Fig1}D). Following \citet{Boromand2019}, we study the relation between the deformation of the droplets and their local packing fraction. To do so, we compute their shape factor $\mathcal{A} = p^2/4 \pi a$ and determine the local packing fraction $\phi_l = a/a_{c}$.  Note that we only consider droplets in the center of the packing, i.e. we exclude those that are partially cut by the edge of the image frame. The shape parameter $\mathcal{A}$ equals 1 for circular disks and is greater than 1 for all nonspherical particles \cite{Boromand2018}.

\subsection{Experimental set-up for emulsion flow}
\par We designed the constriction in a microfluidic channel composed of three main sections (Fig.~\ref{Fig1}): at the entrance, the channel is 50~$\mu$m deep and 200~$\mu$m wide over a 5 mm length, then at the constriction the width is reduced from 200 to 50~$\mu$m over a length of 200~$\mu$m, finally the channel remains 50~$\mu$m wide over a final 5~mm length. The channel is made in polydimethylsiloxane using a negative cast micromachined in a block of PMMA ($50\times50\times5~\mathrm{mm}^{3}$) using a  desktop CNC Mini-Mill machine (Minitech Machinary Corp., USA). After passivating the channel with casein 0.05~mg/ml ($\beta$-casein from bovine milk, Sigma Aldrich) for 20 minutes, the emulsion is flown in the device using a pressure pump (MFCS-8C Fluigent, P = 30 mbar). After droplets fill the constriction area, the pressure is decreased to stop the emulsion flow, and droplets are left to cream in the supply tube overnight, thus compressing the droplets in the microfluidic device in order to reach high values of packing fraction. After this passive compression phase, the emulsion is flown again in the channel at a constant pressure. The flow of the droplets at the constriction is imaged in bright field microscopy with a 10x objective at a frequency of 20 frames per second (fps). 

\subsection{Image analysis of the emulsions flowing in the constriction}
\par To analyse the videos of flowing emulsions, we first threshold the images to subsequently determine the center and perimeter of each droplet in the channel using a custom made Matlab routine. When studying droplet deformation, we only consider the droplets located in the constriction region. We define this area along the channel as a window that includes the 200~$\mu$m of the constriction itself, plus 50~$\mu$m before and after the constriction (Fig.~\ref{Fig1}). To quantify the deformation of each droplet, we use the approach proposed by \citet{Chen2012}. The perimeter of the droplet is discretized at evenly spaced 1024 angles $\theta$ and the deformation $d$ is calculated as a standard deviation of the radii $r(\theta)$ for each of these angles divided by the mean value of $r$:

\begin{equation}
d = \frac{\sqrt{\langle r^2\rangle+\langle r \rangle^2}}{\langle r\rangle}
\end{equation}

We also determine the global packing fraction of the emulsion in each video frame. To this end, we calculate the ratio between the sum of all droplets area and the area of the channel within the window of $200\times200~\mu$m located before the constriction region. Finally, frames are sorted according to the emulsion packing fraction, and the distributions of droplet deformations for each packing fraction are computed.

For rearrangement and flow analysis, the droplets were tracked using a custom Python routine. All droplets are sorted according to the lane they belong to in the channel ahead of the constriction. In our experiments, they are thus sorted into four lanes. The instantaneous velocity of the droplets was computed as the distance travelled between two consecutive frames acquired at a fixed frame rate. The localization of the minima in the instantaneous velocity are then measured for each droplet trajectory and sorted as a function of the original lane the droplet belonged to.

\section{RESULTS}
\subsection{Analysis of static packings}
\par We first study 2D static packings of polydisperse emulsions with two distinct depletion interactions. Using silicon oil droplets stabilized with two different concentrations of SDS (10mM and 45mM) allows us to change the depletion forces between the droplets. In our experiment, the continuous aqueous phase is supplemented in glycerol (40 \% in volume of glycerol). Note that in addition to allowing for a better imaging of the droplets, it also shifts the critical micellar concentration (CMC) of SDS. However, the CMC is only raised from 8mM (in pure water) to about 9mM in our experimental conditions ~\cite{Ruiz2008, Khan2019}, which ensures that the system is still above the CMC under both SDS concentrations and that the surface tension remains the same when the concentration of SDS is increased from 10 to 45 mM. Above the CMC, depletion attraction forces increase linearly with the concentration of micelles \cite{Asakura1958}, which itself grows with increasing concentrations of SDS. Considering the aggregation number of SDS at both SDS concentrations (i.e. the number of SDS molecule per micelle at a given concentration), we estimate that there is approximately 30 times more micelles at 45mM SDS than at 10mM SDS. Depletion forces at 45mM SDS are thus expected to be 30 times larger than at 10mM SDS. 

\begin{figure}
	\centering
		\includegraphics[width=0.45\textwidth]{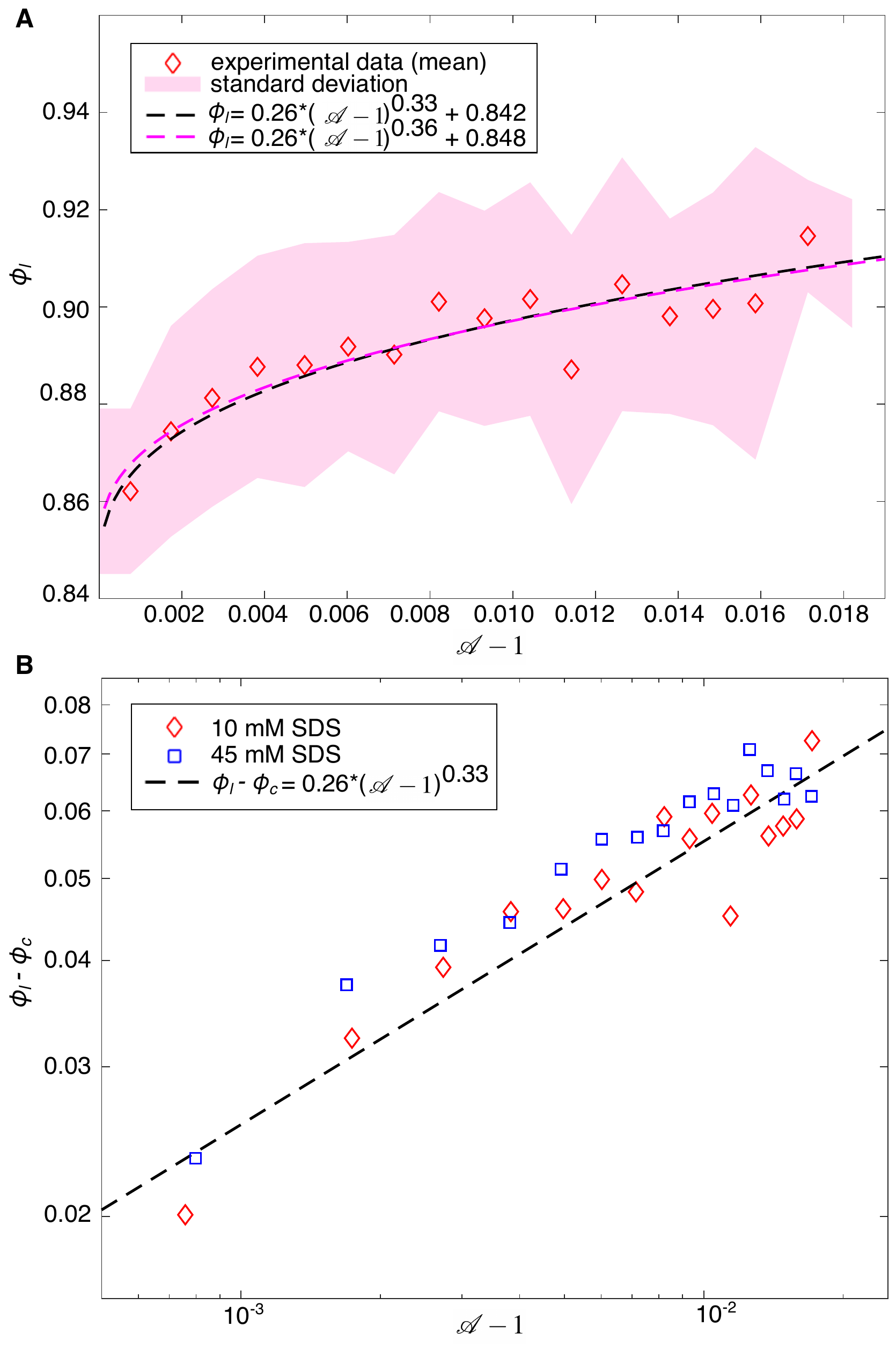}
	\caption{{Analysis of static 2D packings} --- {(A)} $\phi$ versus $\mathcal{A}-1$ for 10mM SDS emulsion. The total number of droplets is N = 1193. The experimental data (red open diamonds) are plotted together with the DP model with the exponent fixed to 1/3 and $\phi_c=0.842$ (black dashed line). The pink dashed line is the best power law fit with the prefactor fixed. {(B)} Log-log plot of $\phi_l-\phi_c$ versus $\mathcal{A}-1$ for 10mM (red open diamonds) and 45mM SDS (blue open squares) emulsions. The data points for the 10mM SDS emulsion are the same as in (A). The total number of droplets for the 45mM SDS emulsion is N = 1735. The DP model is plotted as a black dashed line.}
	\label{Local-contour}
\end{figure}

To study the impact of depletion forces on static 2D packings, we first quantify the deformation of the droplets as a function of their local packing fraction. Recent studies\citep{Boromand2018,Boromand2019} developed a new numerical model to study the structural and mechanical properties of 2D bubbles and emulsions, including at high compressions. In the so-called deformable particle (DP) model, particles deform in response to mechanical constraints to minimize their perimeter while keeping their area fixed. 
This leads to a model of deformable disks with potential energies that includes an energy term associated to the line tension and a penalization energy term quadratic in the change of area of the droplets, thus associated to their compressibility. Further, the deformable particles interact via a purely repulsive potential energy.
Within the framework of this DP model and in our range of deformations, it was predicted that the distance to jamming onset $\phi_l - \phi_c$ scales  with asphericity $\mathcal{A}-1$ as $\phi_l - \phi_c \sim (\mathcal{A}-1)^\omega$ with $\omega \approx 0.3$. 
  
Thus, we measure the asphericity and local volume fraction of each droplet in several images of 2D packings for both 10 and 45mM SDS concentrations (see Fig.~\ref{Fig1}C-D and Materials and Methods). In Fig.~\ref{Local-contour}A, we first plot $\phi_l$ \textit{vs} $\mathcal{A}-1$ for 10mM SDS emulsions in order to compare our data with the theoretical predictions. We then fit our data with the equation $\phi_l = \alpha (\mathcal{A}-1)^{1/3} + \phi_c $ with a fixed $\phi_c=0.842$ as numerically computed in~\citet{Boromand2019}. We find a good agreement between theory and experiments with a prefactor $\alpha=0.26$. Conversely, when we only fix the prefactor $\alpha=0.26$, and keep both $\omega$ and $\phi_c$ as free fitting parameters, we recover $\omega = 0.36\pm0.1$ and $\phi_c=0.848\pm0.02$, which are also very close to the numerically computed ones. For the 45mM SDS emulsion, we also find that $\omega=0.37\pm0.07$ and $\phi_c=0.852\pm0.02$. As shown in Fig.~\ref{Local-contour}B, both SDS concentrations cannot be distinguished. Indeed, the data points corresponding to both depletion forces overlap and are captured by the same scaling that was predicted by the DP model for repulsive discs. This indicates that depletion attraction does not induce any measurable modification in the static packings of droplets. In other words, the interaction energy term that could be included in the DP model does not affect significantly the scaling of  $\phi_l - \phi_c$ \textit{versus} $\mathcal{A}-1$ for depletion induced attractive interactions. 

\begin{figure}[h!]
	\centering
		\includegraphics[width=0.45\textwidth]{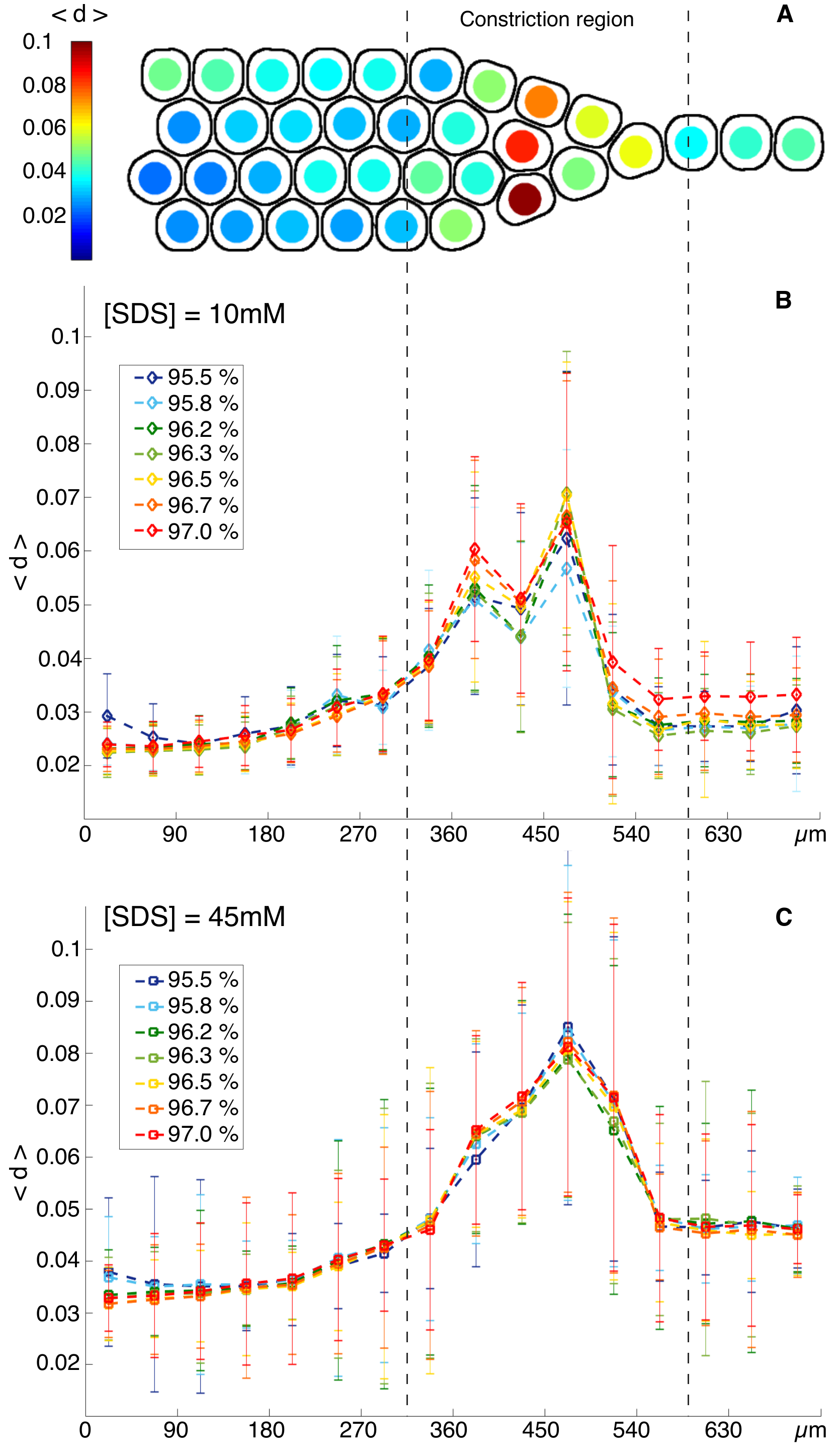}
		\caption{{Analyzing the droplet deformation in the constriction} --- {(A)} Still snapshot of the image analysis in the channel at a given instant for an attractive emulsion ([SDS]=45mM). The color of the droplets codes for their deformation $d$ calculated for their detected contours displayed on the image. {(B-C)} Average deformation of the droplets along the x-axis of the channel for different packing fractions in {(B)} the low attraction case ([SDS]=10mM) and {(C)} high attraction case ([SDS]=45mM). The deformation is averaged in bins that are 45 $\mu$m wide along the x-axis, corresponding to about a droplet diameter. The average deformation peaks inside the area of the constriction for both conditions. The error bars correspond to the standard deviation of the distributions of $d$ calculated for all droplets in all experiments in each bin. The total number of droplets, combining all packing fractions, is N = 27219 for 10mM SDS and N = 91391 for 45 mM SDS.}
	\label{Xspace}
\end{figure}

Despite the fact that static packings cannot be distinguished as a function of depletion forces, we reveal in what follows that significantly distinct behaviors can be evidenced in the context of a dynamic flow. 

\subsection{Emulsion flow in a constriction}
\par In order to study their response under mechanical perturbations, monodisperse emulsions are flown in microfluidic channels exhibiting a single physical constriction (Fig.~\ref{Fig1}). In particular, we use monodisperse droplets whose diameter is comparable to the channel height, constraining the system to a 2D monolayer of droplets. We focus our analysis on the area of the constriction in which droplets have to rearrange and deform in order to go from a large channel into a narrower one. The width of the narrow channel is chosen such that it only allows for the passage of one droplet diameter (Fig.~\ref{Fig1}) in order to maximize the number of rearrangements. 

A typical experiment is carried out in two phases. The channel is first filled with the emulsion using a pressure pump. After a waiting time (see Materials and Methods), the pressure is increased again so that this packed emulsion can flow in the channel. We usually require a typical pressure of the order of 30~mbar to establish a continuous flow. For each experiment, we image the droplets upstream, in order to evaluate their packing fraction, as well as inside the constriction to measure their deformation. We choose to quantify the deformation $d$ of each droplet in the channel through the standard deviation of droplet radii as previously done ~\cite{Chen2015} (see Materials and Methods).

\subsection{Deformation along the channel}
\par We first study the deformation of the droplets inside the channel. To do so, we measure the volume fraction of the emulsion in a window located upstream of the constriction (on the left of the image) and that encompasses 200$\mu$m of the channel length (Fig.~\ref{Xspace}A). We show in Fig.~\ref{Xspace} the average deformation $\left<d\right>$ along the channel for both SDS concentrations. 

The shape of the obtained curves differ for the two SDS concentrations both in the constriction region and in the thinner channel. For 10mM SDS (Fig.~\ref{Xspace}B) the deformation builds up in the constriction to a first maximum average deformation until it is released to a lower value of $\left<d\right>$ at $x \approx 420\mu m$. Then the deformation builds up again to a second maximum and is decreased to a lower deformation. Qualitatively, this behavior can be explained as the signature of a local stress release after a rearrangement. Indeed, \citet{Chen2015} showed that in compressed emulsions, T1 events were immediately followed by a local decrease of deformation inside compressed emulsions. Here the observed peaks are separated by about 40$\mu$m, corresponding to a droplet diameter. This indicates that droplet rearrangements indeed occur at positions that are set by the topology of the packing in the channel~\cite{Gai2016}. 
However, at 45mM SDS (Fig.~\ref{Xspace}C) droplets progressively deform and reach a peak deformation value until they escape it. In that framework, the higher attractive depletion forces could impair the order of rearrangements, explaining why we do not observe a localized deformation release compared to the low attraction case. The plastic response of the attractive emulsion is thus less ordered spatially, which in turn leads to a higher elastic contribution.

The other difference between the two conditions can be observed in the thinner channel region, after the constriction, where droplets enter one by one and release their deformation. In the case of low depletion forces ([SDS]=10mM), droplets relax to a deformation value that is close to the initial one at the entry of the channel $\left(\left<d\right>_{out}-\left<d\right>_{in} \approx 0.0025\right)$. However, with high depletion forces ([SDS]=45mM), droplets relax to a plateau at higher values of deformation than at the entry $\left(\left<d\right>_{out}-\left<d\right>_{in} \approx 0.01\right)$. This impaired relaxation could be a signature of long range effects that could also explain why droplets enter the constriction with a slightly higher value of deformation in the high attraction case. 

\begin{figure}[h]
	\centering
		\includegraphics[width=0.45\textwidth]{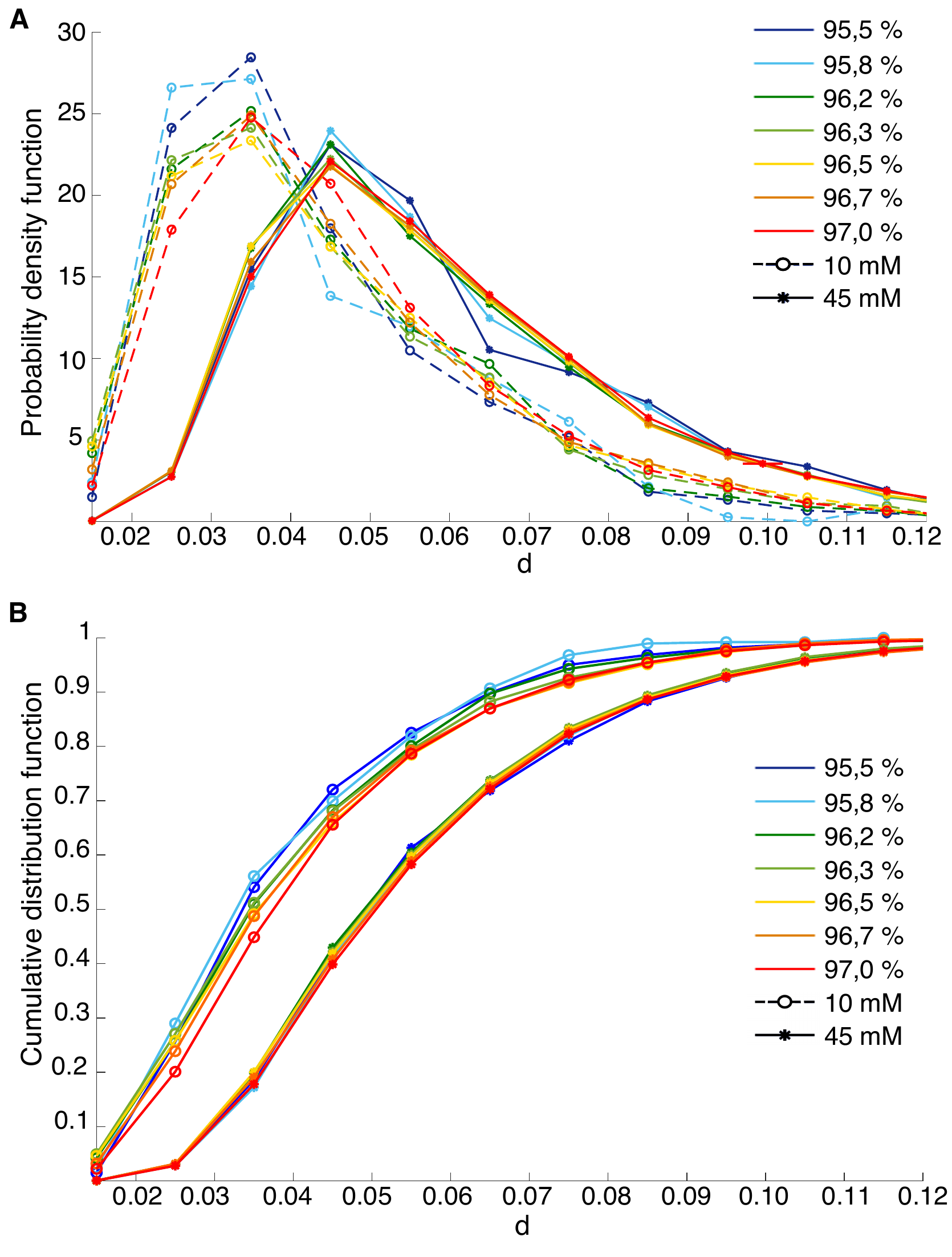}
	\caption{{Statistics of deformation under flow} --- (A) Probability density function of the deformation $d$ calculated in the constriction for different packing fractions in the case of low attraction forces ([SDS] = 10mM, open circles) and  high attraction forces ([SDS] = 45mM, stars). {(B)} Cumulative distributions of the deformation $d$ in the constriction for low attraction forces (open circles) and high depletion forces (stars) for different packing fractions. }
	\label{distributions}
\end{figure}

\subsection{Deformation as a function of packing fraction}
\par To further confirm these observations, we study the distribution of deformation of all droplets at all positions inside the constriction (taken in a window whose length spans 50$\mu$m before and after the constriction -- see Materials and Methods). Since the global volume fraction can evolve over the course of one experiment, we separate each experiment into stacks according to their upstream volume fraction. We then pool together the image sequences corresponding to the same volume fraction throughout all performed experiments, for each concentration. Note that we also checked that the deformation in the constriction does not depend on the instantaneous droplet velocity within the investigated range (from 120 to 360 $\mu m/s$, see ESI$^\dag$).

\par We compare the distributions of the deformations observed for different packing fractions and for each SDS concentration (Fig.~\ref{distributions}). The distributions peak at smaller values of deformation in the low attraction case than in the case of strongly attractive droplets (Fig.~\ref{distributions}A). This shift can also be clearly evidenced when plotting the cumulative distributions for each condition at various volume fractions (see Fig.~\ref{distributions}B). As expected, for low depletion forces (10mM SDS) we find that the distributions exhibit lower values of deformation in all conditions. When attraction is introduced between droplets, all curves are shifted to higher values of deformation. 
In the previous section we showed that depletion alone was not sufficient to induce significant additional deformations in static packings of droplets. The shift  observed in these deformation distributions must thus originate from differences in the local topological changes of the emulsions. Hence, we next examine the spatial localization of rearrangements in the constriction as a function of SDS concentration.

\begin{figure*}
	\centering
		\includegraphics[height=4.5cm]{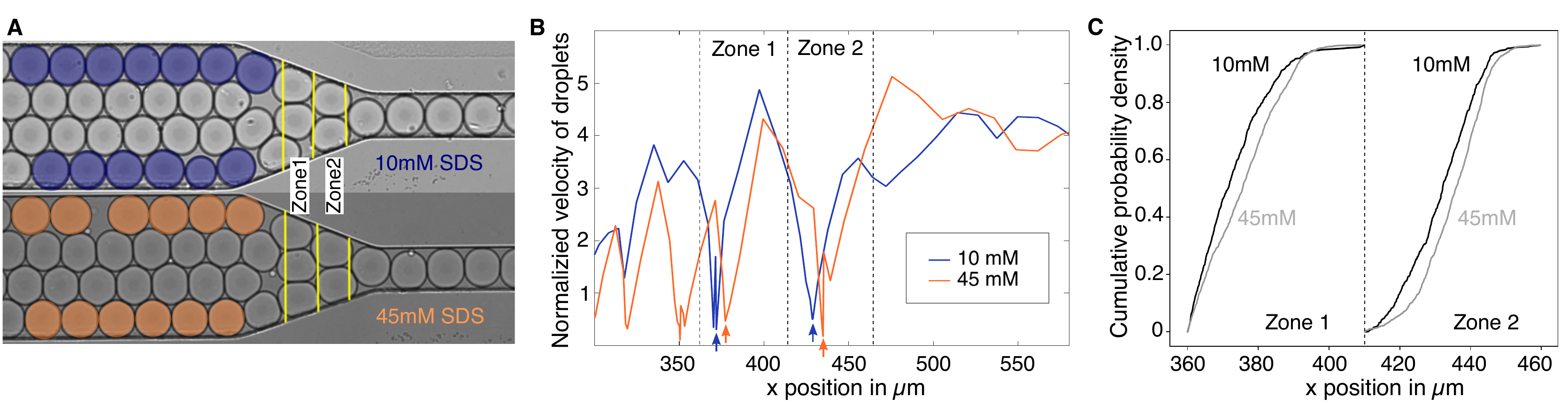}
		\caption{{Rearrangements and velocity distributions in the constriction} --- {(A)} Image of droplets in the constriction for 10mM (top, blue) and 45mM (bottom, orange) SDS. The zones where the number of droplets decreases from 3 to 2 and from 2 to 1 are indicated with yellow lines and referred to as zone 1 and 2 respectively. {(B)} Typical velocity curves of droplets in lines 1 and 4 along the channel axis for 10mM (blue) and 45mM SDS (orange) emulsions. The velocity drops to a minimum value (indicated by an arrow) in the zones 1 and 2 each time the droplets stall before a rearrangement. {(C)} Cumulative distributions of the localization of the minimal velocity for the droplets in all lines for 10mM (black curves) and 45mM SDS (grey curves). In zones 1 and 2, the 45mM SDS droplets slow down further into the constriction, as evidenced by the shift in distributions at 50$\%$ probability, by about 5$\mu$m.}
	\label{Rearrangements}
\end{figure*}

\subsection{Rearrangements and velocity distributions in the constriction}
\par We here test the hypothesis that rearrangements are impaired by the attraction between the droplets, which would in turn force them to deform more to overcome the constriction. To study the rearrangements in the constriction, we simply track the position of the droplets in separate lines of the channel for both conditions (Fig.~\ref{Rearrangements}A). 
Indeed, since the size of the channel as well as the diameter of the droplets are fixed, there are always four lines of droplets flowing in the channel, ahead of the constriction. In this framework the droplets will exchange neighbors to do the necessary rearrangements in given zones of the channel that are defined by geometry. At a point of rearrangement, droplets should thus transiently decrease their speed and subsequently accelerate once the rearrangement is achieved. 
Typical instant velocity profiles are plotted in Fig.~\ref{Rearrangements}B where 2 contiguous zones of rearrangements are highlighted. These zones were chosen as areas where the velocity of the droplets hits a local minimum. In this figure, one can see that the minimum values in zone 1 and 2 seem to be reached further down into the constriction for 45mM SDS than for 10mM SDS (see orange and purple arrows respectively). In order to quantify this observation, we extracted the position on the x-axis of the minimum velocity for each droplet in zones 1 and 2 and plotted their cumulative distributions in Fig.~\ref{Rearrangements}C. The distributions for attractive droplets are both shifted by about 5$\mu$m (measured shift at 50$\%$), indicating that rearrangements are indeed delayed in the channel compared to the low depletion case.

\section{DISCUSSION}
\par Attractive interactions between particles is expected to affect their packing topology as well as their rheological and mechanical response to local mechanical perturbations. Below the jamming transition, previous work showed that attraction induced by depletion forces tuned significantly the structure of 3D packings and could mechanically stabilize them below the isostatic limit~\cite{Jorjadze2011}. 
Above the jamming transition, one expects adhesive forces in packings of deformable spheres to change how droplet deformation and coordination numbers scale with the packing fraction~\cite{Boromand2018,Boromand2019}. To the best of our knowledge, this issue has been addressed neither in theoretical models nor in experimental systems.

In our experimental study, we provide a first step towards the understanding of the mechanical response of adhesive emulsions by introducing attractive interactions induced by depletion between oil droplets. We first evidence that such attraction forces are too low to induce any measurable effect in 2D static packings of droplets. Indeed, for both attraction forces, we recover the scaling laws predicted by ~\citet{Boromand2019} for purely repulsive packings, with a critical packing fraction $\phi_c \approx 0.842$. However, using monodisperse emulsions, we uncovered distinct changes in their elasto-plastic response when the droplets are flown through a 2D physical constriction. The first manifestation of attraction is an increase of the average deformation of the droplets in the constriction. The second one is the delay of topological rearrangements inside the constriction as attraction forces are increased. Depletion forces thus appear adequate to change the elasto-plastic response of emulsions in our system.

Such findings could be relevant for biological tissues in which adhesion controls to a large extent remodelling events that occur on timescales that are beyond those of cytoskeletal activity. In order to isolate the role of adhesion in biological processes, cellular tissues can indeed be mimicked with droplet assemblies connected by specific binders~\cite{Pontani2012, Feng2013, Pontani2016}. Within that framework, emulsions have been shown to exhibit similar mechanical properties and have for this reason been used to measure cellular forces both \textit{in vitro}~\cite{Molino2016} and \textit{in vivo}~\cite{Campas2014, Mongera2018}. This reductionist approach could thus shed light on behavioral transitions in developping tissues upon adhesion modulation and will be the focus of future investigations.\\

\begin{acknowledgments}
The authors thank Yannick Rondelez for providing flow focusing devices for the production of the emulsions, as well as Jacques Fattaccioli for allowing us to use his pressure emulsifier, they also thank Georges Debregeas and Zorana Zeravcic for fruitful discussions and acknowledge financial support from Agence Nationale de la Recherche (BOAT, ANR-17- CE30-0001). 
\end{acknowledgments}

\bibliographystyle{apsrev4-1}

\end{document}